\documentclass[fleqn,usenatbib]{mnras}
\usepackage{float}
\usepackage{newtxtext,newtxmath}
\usepackage[T1]{fontenc}
\usepackage{ae,aecompl,soul,ulem}

\usepackage{graphicx}	% Including figure files
\usepackage{amsmath}	% Advanced maths commands
\usepackage{amssymb}	% Extra maths symbols
\title[Molecular environments of G359.1-0.5]{\vspace{-10mm}\fontsize{20pt}{10pt}\selectfont\textbf{Molecular environments of the supernova remnant G359.1-0.5.} }

\author[L. K. Eppens et al.]{
L. K. Eppens,$^{1,2}$\thanks{E-mail: leppens@iafe.uba.ar}
E. M. Reynoso,$^{1}$
J. Lazendic-Galloway$^{3}$
J. A. Combi$^{2,4}$ and
\newauthor J. F. Albacete Colombo$^{5}$
\\
$^{1}$Instituto de Astronom\'\i a y F\'\i sica del Espacio, CONICET-UBA, Buenos Aires, Argentina\\
$^{2}$Facultad de Ciencias Astron\'omicas y Geof\'\i sicas, Universidad Nacional de La Plata, La Plata, Buenos Aires, Argentina\\
$^{3}$ School of Physics and Astronomy, Monash University, Clayton VIC 3800, Australia.\\
$^{4}$Instituto Argentino de Radioastronom\'\i a, CONICET-CICPBA, Argentina, Buenos Aires, Argentina\\
$^{5}$ Universidad Nacional de R\'\i o Negro, Sede Atl\'antica - CONICET, Viedma CP8500, R\'\i o Negro, Argentina.
}

\date{Accepted 2020 February 14. Received 2020 February 3; in original form 2020 January 7}

\pubyear{2020}
\begin{document}
\label{firstpage}
\pagerange{\pageref{firstpage}--\pageref{lastpage}}
\maketitle

% Abstract of the paper
\begin{center}
\begin{abstract}
{We report new CO observations and a detailed molecular-line study of the mixed morphology (MM) supernova remnant (SNR) G359.1-0.5, which contains six OH (1720~MHz) masers along the radio shell, indicative of shock-cloud interaction. 
Observations of $^{12}$CO and $^{13}$CO J:1--0 lines were performed in a $\sim 38^\prime \times 38^\prime$- area with the on-the-fly technique using the Kit Peak 12 Meter telescope. The molecular study has revealed the existence of a few clumps with densities $\sim 10^3$~cm$^{-3}$ compatible in velocity and position with the OH (1720 MHz) masers. These clumps, in turn, appear to be part of a larger, elongated molecular structure $\sim 34^\prime$ long extending between $-12.48$ and +1.83 km s$^{-1}$, adjacent to the western edge of the radio shell. According to the densities and relative position with respect to the masers, we conclude that the CO clouds depict unshocked gas, as observed in other remnants with OH (1720 MHz) masers. In addition, we investigated the distribution of the molecular gas towards the adjacent $\gamma$-ray source HESS J1745-303 but could not find any morphological correlation between the $\gamma$-rays and the CO emission at any velocity in this region.
}
\end{abstract}
 \end{center}
 
% Select between one and six entries from the list of approved keywords.
% Don't make up new ones.
%%%%%%%%%%%%%%%%%%%%%%%%%%%%%%%%%%%%%%%%%%%%%%%%%%

%%%%%%%%%%%%%%%%% BODY OF PAPER %%%%%%%%%%%%%%%%%%
\section{Introduction}\label{Intro}

Core-collapse supernova (SN) explosions are the final stage in the evolution of high mass stars ($\geq$8~M$_{\rm \odot}$). These events produce a diffuse component and a compact object (neutron star or black hole). The diffuse component, called supernova remnant (SNR), results from the combination of the leftover stellar material and the interstellar gas swept up by the shock wave driven by the SN outburst. The expanding shock wave creates a shell morphology that radiates non-thermal emission in radio band (sometimes observed in X-rays as well), created form electrons accelerated in magnetic field form the compressed ISM. At the same time, the expanding shock carries ejected hot stellar material and heats the swept up ISM, and these two components can be detected in the X-ray band as thermal emission. However, there is a fraction of SNRs for which the X-ray emission, rather than forming a shell, fills the centre of the radio remnants, creating a separate morphological group named mixed- morphology (MM) SNRs \citep{Rho1998}. The galactic source G359.1-0.5 belongs to this group of remnants.

 %%%%%%%%%%%%%%%%%%%%% FIGURE 1 %%%%%%%%%%%%%%%%%%%%%%%%%%%%%%%%%%%
\begin{figure*}
\centering
    \includegraphics[width=0.9\textwidth]{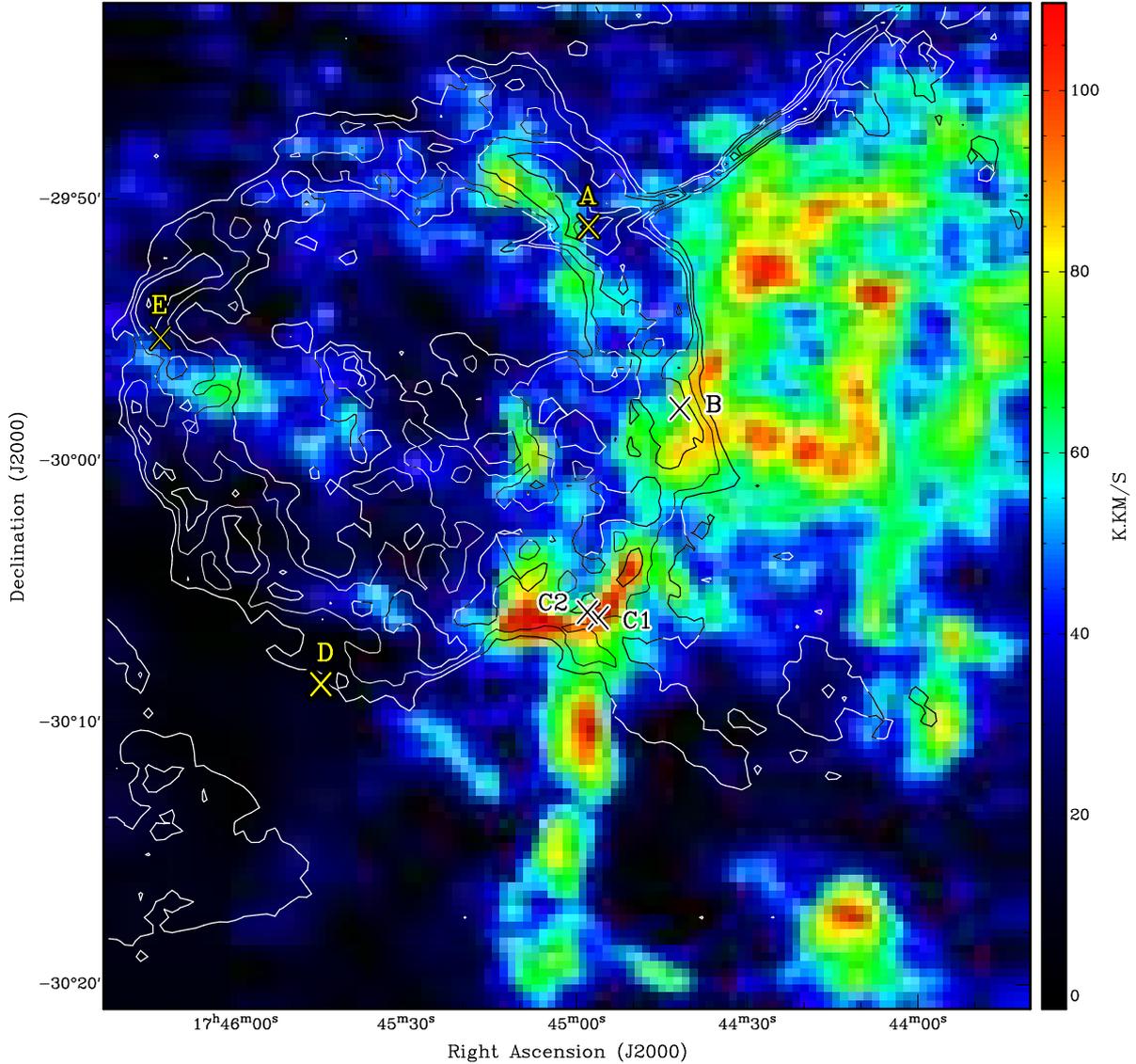}
    \caption{
    Distribution of $^{12}$CO emission integrated between -12.48 and +1.83~km~s$^{-1}$. The beam size is  53\hbox{$.\!\!{}^{\prime\prime}$}6~$\times$~53\hbox{$.\!\!{}^{\prime\prime}$}6. The color scale is indicated at the right in units of K~km~s$^{-1}$. Contours (white or black depending on background) at 0.003, 0.005, 0.008, 0.014, 0.019, and 0.025~mJy~beam$^{-1}$ represent the radio emission at 1.5~GHz. The crosses indicate the location of the OH (1720~MHz) masers reported in \citet{YusefZadeh1995}, and are labeled following the same nomenclature.}
    \label{fig:filam}
    \vspace{5pt}
\end{figure*}
%%%%%%%%%%%%%%%%%%%%%%%%%%%%%%%%%%%%%%%%%%%%%%%%%%%%%%%%%%%%%%
G359.1-0.5 was initially detected by \citet{Altenhoff1979} in a Galactic Plane survey at 4.9~GHz. A flux density of $\sim$13~Jy was estimated. Subsequent observations at 2.7~GHz and 4.8~GHz \citep{Reich1984} classified it as a shell type SNR based on the non-thermal spectrum ($\alpha$ = $-0.37$) and polarized emission (11.5~$\%$). 

There are three X-ray studies that reveal the thermal nature of the filling plasma inside G359.1-0.5. In the first one, \cite{Egger1998} identified diffusive X-rays with the %R$\ddot{o}$ntgensatellit
R\"ontgensatellit
(ROSAT) All Sky Survey \citep[RASS;][]{Truemper1992,Truemper1993} observations, classifying this remnant in the MM group. They fitted the data with a two temperature thermal plasma with the cooler plasma abundant in Silicon (Si), while the hotter component was extremely over abundant in Sulfur (S). Those results implied that the thermal X-rays emission originated in the ejecta rather than swept up ISM.
The second analysis was performed with the Advanced Satellite for Cosmology and Astrophysics (ASCA) by \cite{Bamba2000}, where the authors recognized prominent K$\alpha$ lines of Si and S in the spectrum which were consistent with He-like and H-like ions respectively. However, the third X-ray study based on Suzaku observations \citep{Ohnishi2011} disagreed with the two-component model and proposed an over-ionized state of the inner plasma. \cite{Ohnishi2011} suggested that the ions are cooling faster than electrons, supposedly as a consequence of interaction with a denser environment.

%%%%%%%%%%%%%%%%%%% MASERS DATA TABLE %%%%%%%%%%%%%%%%%%%%%%%%%%%%%%%%%%%
\begin{table}
  \begin{center}
%  \{Tabla de los par\'ametros de la emisi\'on m\'aser de OH}
%  \multicolumn{7}{|c|}{Emisi\'on m\'aser}
%  \hline
   \caption{OH (1720~MHz) masers}
    \label{tab:OH}
   \begin{tabular}{p{0.9cm} p{1.5cm} p{1.6cm} p{1.3cm} p{1.0cm}}
   \hline
    maser&RA(J2000)&Dec.(J2000)&V$_{\rm LSR}$~[km/s]&$S$~[mJy]\\
   \hline
   \ \textbf{A}&17:44:58.06&-29:51:03.7&-4.47&761\\
   \ \textbf{B}&17:44:42.026&-29:58:00.49&-2.9&59.5\\
   \ \textbf{C$_1$}&17:44:56.27&-30:06:00.8&-3.19&215\\
   \ \textbf{C$_2$}&17:44:58.19&-30:05:47.87&-5.57&732\\
   \ \textbf{D}&17:45:44.92&-30:08:33.4&-5.28&639\\
   \ \textbf{E}&17:46:12.96&-29:55:19.0&-5.08&93\\
   \hline
   \end{tabular}
   \end{center}
   \vspace{2pt}
   {\bf {Note.}} Names and parameters of the OH (1720~MHz) masers are taken from \citet{YusefZadeh1995}$^3$.\\
\end{table}
%%%%%%%%%%%%%%%%%%%%%%%%%%%%%%%%%%%%%%%%%%%%%%%%%%%%%%%%%%%%%%%%%%%%%%%
G359.1-0.5 belongs to a group of SNRs that have associated OH (1720~MHz) masers \citep{Frail1998,Green1997,Koralesky1998}. These particular masers are tracers of interaction between SNRs and molecular gas, since they are excited when the shock front hits the surrounding molecular material \citep{Elitzur1976}. \citet{YusefZadeh1995} found six masers strikingly distributed along the edge of G359.1-0.5 with ve\-lo\-ci\-ties between $-5.57 < v < -2.9$ km~s$^{-1}$. They discarded that the 1720 MHz line emission had origin in spiral arms foreground dark clouds \citep[e.g.][]{Turner1982} which amplify background radiation, since no continuum emission is detected at the locations of masers B and D, while regions of brighter continuum emission over the shell or even beyond do not coincide with any 1720 MHz line emission spot. Also, the fact that no masers are found in the interior but all of them are on the outer shell, together with an extended 1720 MHz component that closely follows the edge of the remnant (see Sect. \ref{sec:coyrsn}), strongly suggests that shocks are involved. Based on all this evidence, they proposed that the OH (1720 MHz) masers are being generated behind the shock front and, therefore, are associated with the SNR.

\cite{Uchida1992b} inspected the molecular distribution towards G359.1-0.5 in the $^{12}$CO J=1-0 line and found a nearly continuous ring of molecular gas ($-190$ <\textit{v}< $-60$~km~s$^{-1}$) coincident with the radio shell. A follow-up study of the H{\sc i} line using VLA data \citep{Uchida1992a} shows a counterpart of the CO ring between -190 and -75~km~s$^{-1}$. A simultaneous H{\sc i} absorption study showed that there were H{\sc i} features at -60 and -135~km~s$^{-1}$ towards G359.1-0.5. However, the subsequent discovery of the OH (1720~MHz) masers at a significantly different velocity casts doubts on the connection between this annular cloud and the SNR. \citet{Lazendic2002} obtained new high-resolution molecular data focusing on a smaller field towards maser A. They found a shocked cloud coincident in position and velocity with the maser.

An extended unidentified $\gamma$-ray source, HESS J1745-303, was also detected in the vicinity of G359.1-0.5 \citep{Aharonian2006}. A GeV counterpart to this TeV source was detected with the Fermi Large Area Telescope \citep[LAT; ][]{Hui2011}. The nature of HESS J1745-303 is still uncertain and its association with G359.1-0.5 remains a matter of debate.

This paper reports original 12 Meter Radio Telescope observations in the $^{12}$CO and $^{13}$CO J:1--0 lines with an improved resolution as compared to previous surveys covering the full extent of G359.1-0.5 and surroundings. We aim to re-visit the molecular gas distribution around the SNR and confirm or reject the previous claims for CO structures related to it.
The outline of the paper is as follows: in Sect.~\ref{sec:obs} we describe the molecular line observations and data reduction. The results of the molecular analysis is presented in Sect.~\ref{sec:radio}. In Sect.~\ref{sec:discuss}, we discuss the implications of our results, including the possible connection between the molecular gas distribution and the origin of the $\gamma$-rays emission. Finally, we summarize the main conclusions in Sect.~\ref{sec:conclu}.

\section{Observations and data reduction}
\label{sec:obs} 

Observations were performed on February 2002 with the Arizona Radio Observatory (ARO) 12M Radio Telescope located at Kitt Peak, Arizona. The observed field, covering $\sim 38^\prime \times 38^\prime$, was scanned applying the on-the-fly (OTF) technique, efficient for imaging extended regions. Two spectrometers centered at $-5.0$~km~s$^{-1}$ were used, both having 256 channels but with different resolutions, of 250 and 500 kHz (0.65 and 1.3~km~s$^{-1}$) for the $^{12}$CO line and 100 and 500 kHz (0.27 and 1.35~km~s$^{-1}$) for the $^{13}$CO. The $^{12}$CO line (centered at 115.271204~GHz) was observed during 4.5 hours on February 23rd, while the $^{13}$CO (centered at 110.201353~GHz) was observed for 3.25 hours on February 24th. 

Data processing was carried out using the AIPS package. A set of line free channels was selected for baselines subtraction, which was applied to the whole map through the task SDLSF. The line-free channels selection was aided with an H{\sc i} spectrum obtained from the Leiden-Argentina-Bonn Survey \citep{Kalberla2005}. Data editing was performed with WIPER and corrupted observations were removed. The data cubes were constructed with SDGRD. A few spread channels needed to be fully removed, and were replaced by an average between the preceding and subsequent channels. The spatial resolution is $53\farcs 6 \times 53\farcs 6$ and $56\farcs 0 \times 56\farcs 0$ for the $^{12}$CO and $^{13}$CO respectively.

\section{Results}
\label{sec:radio}
%%%%%%%%%%%%%%%%%%%%%%%%%%% CO RESULTS TABLE %%%%%%%%%%%%%%%%%%%%%%%%%%%%%%%%%%%%%%%%%%%%%
\begin{table*}
       \caption{Physical parameters of CO clumps}
        \label{tab:resultados}
       \begin{center}
       \begin{tabular}{p{3.0cm} p{1.6cm} p{1.4cm} p{1.4cm} p{1.4cm}  p{1.4cm} p{1.4cm} }
       \hline
       \ Parameters&\textbf{a$_1$}&\textbf{a$_2$}&\textbf{b$_1$}&\textbf{c$_1$}&\textbf{c$_2$}&\textbf{e$_1$}\\
       \hline
       \ RA(J2000) ($^{\rm h} ~ ^{\rm m} ~ ^{\rm s}$)&17~45~3.1&17~45~1.7&17~44~38.1&17~44~49.1&17~44~51.9&17~46~2.6\\
       \ Decl.(J2000) ($^{\circ}~^\prime~ ^{\prime\prime}$)&-29~50~10.01&-29~51~3.9&-29~57~3.5&-30~07~15.8&-30~05~9.8&-29~57~21.9\\
       \ \textit{v}$_i$,\textit{v}$_f$($^{12}$CO)~(km~s$^{-1}$)&0.85,-8.25&0.85,-8.25&0.53,-5.98&-1.42,-4.67&-3.05,-8.25&-2.07,-7.93\\
       \ \textit{v}$_i$,\textit{v}$_f$($^{13}$CO)~(km~s$^{-1}$)&-&-&0.58,-5.45&-2.42,-3.5&-&-2.14,-7.86\\
       \ T$_{\rm peak}$($^{12}$CO)~(K)&10.5&9.8&11.3&13.3&16.4&10.7\\
       \ T$_{\rm peak}$($^{13}$CO)~(K)&-&-&3.9&4.6&-&4.3\\
       \ T$_{\rm exc}$~(K)&-&-&14.7&16.7&-&14.1\\
       \ $\tau$($^{13}{\rm CO}$)&-&-&0.43&0.42&-&0.51\\
       \ \textit{N}($^{13}{\rm CO}$)~(10$^{15}$~cm$^{-2}$)&-&-&32.6&6.8&-&31.5\\
       \ $N_{\rm H_2}$~(10$^{21}$~cm$^{-2}$)&8.3&5.7&16.3&3.4&10.2&15.7\\
       \ $M_{\rm H_2}$~($(\frac{d}{3.7})^2$M$_{\odot}$)&$\sim$3080&$\sim$190&$\sim$955&$\sim$145&$\sim$690&$\sim$2340\\
       \ $n_{\rm H_2}$($\frac{3.7}{d} 10^3$~cm$^{-3}$)&$\sim$1.8&$\sim$2.1&$\sim$13.8&$\sim$1.1&$\sim$5.6&$\sim$4.5\\
       \hline
       \end{tabular}
       \end{center}
     \vspace{2pt}
       {\bf Note.} Physical parameters for CO clumps associated with the OH (1720~MHz) masers. The velocities \textit{v}$_i$ and \textit{v}$_f$, in units of km~s$^{-1}$, indicate the integration range. The mass $M_{\rm H_2}$ and the volume density $n_{\rm H_2}$ are scaled with the distance in units of kpc.
       
\end{table*}
%%%%%%%%%%%%%%%%%%%%%%%%%%%%%%%%%%%%%%%%%%%%%%%%%%%%%%%%%%%%%%%%%%%%%%%%%%%%%%%

As mentioned above, G359.1-0.5 is one of the SNRs for which OH (1720~MHz) masers were detected. The physical conditions for the collisional excitation of these masers are moderate temperatures (50<T<125~K) with densities 10$^{4} \leq n_{\rm H} \leq 5 \times 10^{5}$ cm$^{-3}$ and OH column densities of $10^{16} -10^{17}$ cm$^{-2}$ \citep{Lockett1999}. \cite{Frail1996} noted that masers are more easily detected when the acceleration of the molecular gas produced by the SNR shock is transverse to the line of sight. Hence, the velocity of the OH (1720~MHz) masers is coincident with the systematic velocity of the pre-shock molecular gas, with a very low dispersion around this value.  

In Table~\ref{tab:OH} we reproduce the masers parameters as obtained from \cite{YusefZadeh1995}. The first column indicates the name of each maser following the same nomenclature as in their paper. The next two columns list the coordinates converted to J2000.0, including the correction in the declination for maser D \citep{Qiao2018}\footnote{\cite{Qiao2018} detected a typo error in \cite{YusefZadeh1995}. We show the corrected value.}. These positions are indicated by crosses in Fig.~\ref{fig:filam} (see below). The fourth column presents the radial velocity for each maser in km~s$^{-1}$ and the last one shows the flux density in units of mJy. 

To compare the CO distribution with G359.1-0.5, we used a radio continuum image obtained with the Australian Telescope Compact Array (ATCA) at 1.5~GHz \citep{Yusef2004}. This image has a resolution of 20\hbox{$.\!\!{}^{\prime\prime}$}87~$\times$18\hbox{$.\!\!{}^{\prime\prime}$}20, P.A.=-180\hbox{$.\!\!{}^\circ$}5. We inspected the CO data cubes throughout the whole velocity range in order to identify molecular gas features that could be coincident in position and velocity with the OH (1720~MHz) masers. For most of the masers, we found small clouds associated with them, which are embedded in an extended structure extending in velocity from $-12.48$~km~s$^{-1}$ to +1.86~km~s$^{-1}$. This feature can be seen in Fig.~\ref{fig:filam} at the west of G359.1-0.5, where the $^{12}$CO emission is integrated over this velocity interval. In the region overlapping the continuum emission, this CO structure follows the curvature of the radio shell and continues to the south as a tail about 10$^{\prime}$ long. In what follows we will describe the individual structures likely to be associated with each OH (1720~MHz) maser. 

\begin{figure*}
\centering
    \includegraphics[width=0.88\textwidth]{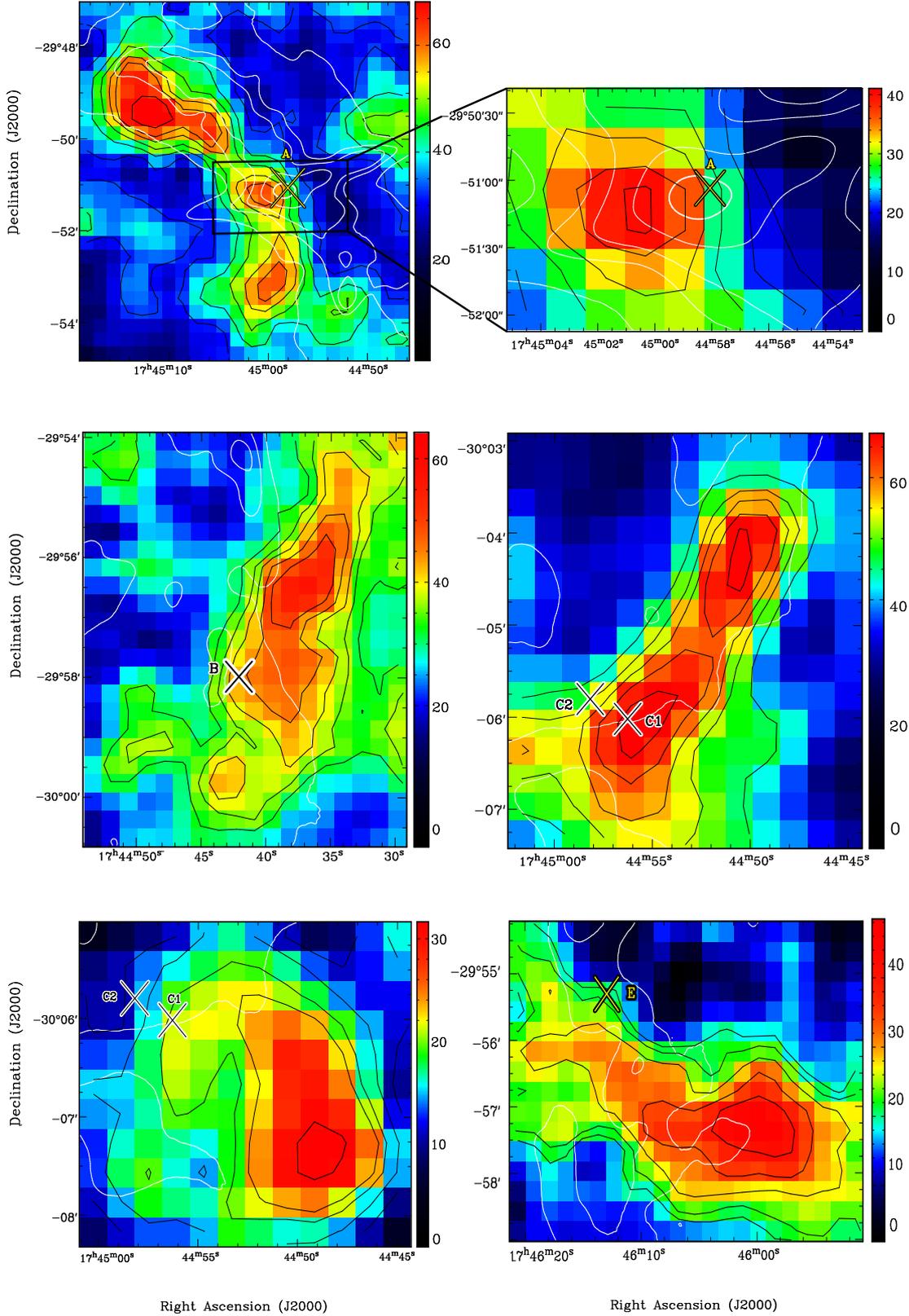}
    \caption{Distribution of $^{12}$CO clouds probably associated with each OH (1720~MHz) maser, whose positions are indicated with crosses. The beam size is 53\hbox{$.\!\!{}^{\prime\prime}$}6~$\times$~53\hbox{$.\!\!{}^{\prime\prime}$}6. The color scale is indicated at the right in units of K~km~s$^{-1}$, and black contours are overlaid to highlight the detected structures. White contours at 0.003, 0.005, 0.008, 0.014, 0.019 and 0.025~mJy~beam$^{−1}$ are used to plot the radio emission at 1.5~GHz.  
    The integration velocity ranges and black contour levels are as follows: 
    Upper left panel: from $-8.25$ to +0.85~km~s$^{-1}$, levels 32, 41, 47.5, 54.3, 59 and 69~K~km~s$^{-1}$. Upper right panel: close-up towards maser A at the same velocities. Middle left panel: from $-5.98$ to 0.53~km~s$^{-1}$, levels 32, 36, 39.6, 44.2 and 49~K~km~s$^{-1}$. Middle right panel: from 3.05 to $-8.25$~km~s$^{-1}$, levels 45, 52, 56.5, 61.5 and 67~K~km~s$^{-1}$. Lower left panel: from -3.5 to $-2.42$~km~s$^{-1}$, levels 11, 14.8, 19.2, 22.1 and 31.8~K~km~s$^{-1}$. Lower right panel: from $-7.93$ to $-2.07$~km~s$^{-1}$, levels 19, 22, 27, 33.8 and 41~K~km~s$^{-1}$.}
    \label{fig:grumos}
    \vspace{5pt}
\end{figure*}

In Fig.~\ref{fig:grumos} we show in the upper left panel the $^{12}$CO cloud, hereafter ``a$_1$'', correlated with maser A. This arc-shaped structure centered at $-4.35$~km~s$^{-1}$ consists of three clumps. The OH (1720~MHz) maser is located at the edge of the middle clump (see upper right panel in Fig.~\ref{fig:grumos}) , which will be referred to as ``a$_2$'' and is centered at R.A.(J2000)=17$^{\mathrm h}$45$^{\mathrm m}$01\hbox{$.\!\!{}^{\rm s}$}7, Decl.(J2000)=$-29^{\circ}$51\hbox{$^{\prime}$}3\hbox{$.\!\!{}^{\prime\prime}$}9 

The rest of the clouds identified close to the other five masers are shown in Fig.~\ref{fig:grumos}.
The second row left panel displays the $^{12}$CO distribution around maser B integrated from $-5.92$ to 0.58~km~s$^{-1}$. This figure shows an extended clump, hereafter ``b$_1$'', centered at R.A.(J2000)=17$^{\mathrm h}$44$^{\mathrm m}$38\hbox{$.\!\!{}^{\rm s}$}1, Decl.(J2000)=$-29^{\circ}$57\hbox{$^{\prime}$}3\hbox{$.\!\!{}^{\prime\prime}$}5, which has a $^{13}$CO counterpart.

The masers C$_1$ and C$_2$ are positionally very close to one another, but differ in velocity by more than 2~km~s$^{-1}$. We searched for a CO structure compatible with C$_1$ and found the cloud enclosed by the contour level of 22.1~K~km~s$^{-1}$ shown in the lower left panel in Fig.~\ref{fig:grumos}. This structure, hereafter ``c$_1$'', is observed both in $^{12}$CO and $^{13}$CO around $\sim -3$ km~s$^{-1}$ and consists of a $\sim$2$^{\prime}$ long concentration extending N-S with an asymmetric intensity distribution, where the peak is at the south, at R.A.(J2000)=17$^{\mathrm h}$44$^{\mathrm m}$49\hbox{$.\!\!{}^{\rm s}$}1, Decl.(J2000)=$-30^{\circ}$07\hbox{$^{\prime}$}15\hbox{$.\!\!{}^{\prime\prime}$}8. The maser lies somewhat afar from c$_1$ at the edge of a weak $\sim$1$^{\prime}$ protrusion with no $^{13}$CO counterpart that extends to the SE from the northern extreme. 

In order to identify the molecular gas associated with maser C$_2$, we inspected the CO cubes at velocities close to $-5.6$~km~s$^{-1}$ and found the $^{12}$CO structure displayed in the right panel at the second row in Fig.~\ref{fig:grumos}. This elongated cloud, only detected in $^{12}$CO, has two peaks at the extremes. The peak closest to maser C$_2$ is centered at R.A.(J2000)=17$^{\mathrm h}$44$^{\mathrm m}$51\hbox{$.\!\!{}^{\rm s}$}9, Decl.(J2000)=$-30^{\circ}$05\hbox{$^{\prime}$}9\hbox{$.\!\!{}^{\prime\prime}$}8 and we will name it ``c$_2$''.

The lower right panel in Fig.~\ref{fig:grumos} shows a $^{12}$CO concentration integrated from $-2.07$ to $-7.93$~km~s$^{-1}$, being maser E located outside the northern limit of the cloud. This structure extends along $\sim$7$^{\prime}$ and has a counterpart in the $^{13}$CO line. The cloud peak is centered at R.A.(J2000)=17$^{\mathrm h}$46$^{\mathrm m}$2\hbox{$.\!\!{}^{\rm s}$}6, Decl.(J2000)=$-29^{\circ}$ 57\hbox{$^{\prime}$}21\hbox{$.\!\!{}^{\prime\prime}$}9 and we will call it ``e$_1$''. Maser D is one of the most intense of the OH (1720~MHz) masers reported by \citet{YusefZadeh1995}. However, we did not find any molecular emission in its boundaries throughout the whole velocity range observed. 
%%%%%%%%%%%%%%%%%%%%%%%%%%%%%%%%% FIGURE 2 %%%%%%%%%%%%%%%%%%%%%%%%%%%%%%%%%%%%%%
%%%%%%%%%%%%%%%%%%%%%%%%%%%%%%%%%%%%%%%%%%%%%%%%%%%%%%%%%%%%%%%%%%%%%%%%%

To estimate the masses of the clumps mentioned above, we assumed local thermodynamic equilibrium. We integrated the H$_2$ column density (\textit{N}$_{\rm H_2}$) over the solid angle ($\Omega$) subtended by each CO emission feature and assumed that the molecular gas is at the same distance as G359.1-0.5. The distance to the SNR is not well constrained. While Galactic rotation models are not accurate towards this direction of the Galaxy, there are suggestions that the SNR is at the distance of the Galactic Center, 8.5~kpc, based on the comparison with X-ray sources with similar absorbing column densities \citep[e.g.][]{Ohnishi2011}. \citet{YusefZadeh1995} sets this distance as an upper limit, since otherwise the SNR diameter would be unrealistically large. We will use the reddening-distance model of \citet{chen+99} to obtain a new, independent estimation. The details on the application of this method are as in \citet{Reynoso2017}. The reddening E(B-V) is derived from the hydrogen column density $N_{\rm H}$ obtained by measuring the X-ray extinction, estimated to be $2 \times 10^{22}$~cm$^{-2}$ \citep{Ohnishi2011}. This gives E(B-V)=2.9~mag. The total reddening produced by the Galactic Plane in the direction of G359.1-0.5 up to the edge of the Galaxy is E(B-V)$_\infty = 13.3 \pm 1.3$~mag \citep{Schlafly2011}\footnote{An on-line tool for obtaining the dust reddening for a given line of sight is supplied in the web site https://irsa.ipac.caltech.edu/applications/DUST/ .}. Assuming a scaleheight absorbing dust of $117.7 \pm 4.7$~pc for the Galactic Plane \citep{Kos2014} and a distance of $19.6 \pm 2.1$~pc from the Galactic Plane to the Sun \citep{Reed06}, the model of \citet{chen+99} produces a distance of $3.7 \pm 0.8$~kpc for G359.1-0.5. At this distance, the SNR diameter is estimated to be $\sim 25$~pc, much more reliable that the 60~pc diameter that the remnant would have at the Galactic Center distance. Nevertheless, we express
 our final results in terms of $d$ (see Table~\ref{tab:resultados}) to facilitate comparison with the results when other distances are assumed.

The $^{13}$CO column density can be computed using 
\begin{equation}
\textit{N}~(^{13}{\rm CO})~ = ~2.42\times10^{14}~\tau(^{13}{\rm CO})~\frac{\Delta v~T_{\rm exc}}{1-\rm e^{-5.29/T_{\rm exc}}}\ [{\rm cm}^{-2}],
\end{equation}
where $\Delta v$ is the FWHM line width in units of km~s$^{-1}$, T$_{\rm exc}$ is the excitation temperature of the J=1$\to$0 transition of the $^{13}$CO molecule in units of K and $\tau$($^{13}$CO) is the optical depth given by
\begin{equation}
  \tau~(^{13}{\rm CO})~ = ~ - \ln \left(1-\frac{T_{\rm peak}~(^{13}{\rm CO})}{5.29~[ (\rm e^{{5.29}/{T_{\rm exc}}}-1)^{-1} - 0.164]}\right).
\end{equation}
 
The excitation temperature was estimated by observing the peak temperature of the optically thick CO (J=1$\to$0) emission, and following the equation:
\begin{equation}
 T_{\rm exc}~ = ~\frac{5.53}{\ln[1+5.53/(T_{\rm peak}(^{12}{\rm CO})+0.819)]}\ [\rm K].
\end{equation}
The \textit{N}$_{\rm H_2}$ of the clumps is directly derived from the fractional abundance \textit{N}$_{\rm H_2}$/\textit{N}($^{13} \rm CO$)=5$\times 10^5$ \citep{Dickman1978}.

Then, the molecular mass of H$_2$ can be calculated using the formula: 
\begin{equation}
  \textit{M}_{\rm H_2}~ = ~ \frac{2.72~m_{\rm H}}{m_{\rm \odot}}~\textit{N}_{\rm H_2}~\Omega(^{12}{\rm CO})~d^2 \ [\rm {M_\odot}],
\end{equation}
where we assumed a mean molecular weight per H$_2$ molecule of 2.72~m$_{\rm H}$ \citep{Allen1973}, $N_{\rm H_2}$ is in units of cm$^{-2}$, $\Omega$($^{12}$CO) is given in units of str and $d$ is the distance in units of cm. 

For those clouds where only $^{12}$CO emission was detected, we calculate $N_{\rm H_2}$ using the relationship X=$N_{\rm H_2}$/W$_{\rm CO}$, where W$_{\rm CO}$ is the integrated CO line intensity in units of K~km~s$^{-1}$ and X = 1.6$\times$10$^{20}$~cm$^{-2}$/K~km~s$^{-1}$ \citep{Hunter1997}.

Table~\ref{tab:resultados} shows the physical parameters obtained for the CO clumps associated with the OH (1720~MHz) masers, where the names of the clumps are given at the top from column two to column seven. The clumps central position (R.A., Decl.) are given in the first two rows. The next two rows show the velocity range where every molecular structure was detected and specify if we found $^{12}$CO and/or $^{13}$CO emission. Rows five and six present the $^{12}$CO and, if applicable, $^{13}$CO T$_{\rm peak}$ in units of K obtained from the observations, and row seven lists T$_{\rm exc}$ in units of K. The calculated parameters $\tau$($^{13}$CO) and \textit{N}($^{13} \rm CO$) in units of cm$^{-2}$ are given in rows eight and nine. The next row shows \textit{N}$_{\rm H_2}$ in units of cm$^{-2}$. Finally, the last two rows display the total mass \textit{M}$_{\rm H_2}$ in solar masses and the volume density $n_{\rm H_2}$ in units of cm$^{-3}$ for each clump; both parameters are scaled with the distance in units of kpc.

\section{Discussion}
\label{sec:discuss}
\subsection{CO distribution at the maser velocities}
\label{sec:coyrsn}

We surveyed the molecular gas towards the OH (1720 MHz) masers and found small cloudlets coincident with most of them both in position and velocity. Only two of the clumps have no $^{13}$CO counterpart, probably due to the low density of
$^{12}$CO which, given that the abundance ratio $N^{12}/N^{13} \sim$20 \citep{Wilson1994}, translates into an even lower $^{13}$CO density, below the detection limit of our data. The cloudlets are, in turn, embedded within an extended $^{12}$CO feature to the West of the SNR radio shell contained in the
velocity range $-12.48 < v < 1.86$ km~s$^{-1}$. In some cases, the gas distribution follows closely the curvature of the SNR. For example, at
the location of the masers C$_1$ and C$_2$, Fig.~\ref{fig:filam} reveals a bright CO arc that strikingly matches the outer radio contour. A similar trend is observed around masers A and B in Fig.~\ref{fig:grumos}. We interpret such remarkable morphological coincidence as an indication of a physical association between the SNR and the molecular gas. Maser E, in contrast, not only does not lie over or very close to any CO cloud but also the nearest cloud cuts the radio contours from east to west rather than following the edge (see Fig. \ref{fig:filam}). Therefore, we propose that this clump is merely a molecular gas feature not related to the maser or to the remnant itself.

The average difference between the velocities of the OH (1720 MHz) masers and those of the corresponding CO clumps is $\sim$0.12~km~s$^{-1}$, much smaller than the spectral resolution of our data. We can therefore conclude that the clumps are at the same velocity as the OH (1720~MHZ) masers. A major obstacle found by \citet{YusefZadeh1995} to support that the OH (1720) MHz masers originate in G359.1-0.5 was the inconsistency between their velocities and the ambient molecular gas, since they assumed that the molecular cloud associated with the SNR was the CO ring reported by \citet{Uchida1992b} beyond $-60$ km s$^{-1}$. Our results sort out this problem.

As regards to the relative position, the distance between each maser and the peak in the CO brightness distribution is typically $\sim$2-3~pc, reaching maser C$_1$ a distance of 5~pc to the emission peak of the clump c$_1$. Similar distances were reported between OH (1720 MHz) masers at other SNRs and their corresponding molecular clumps \citep[e.g.][]{Frail1998, Reynoso2000}. The fact that the masers are near but not at exactly the same position of the CO peaks indicate that CO lines at low rotational transitions do not trace regions with density high enough to excite OH (1720 MHz) maser emission. Such density was estimated to be $n_{\rm H_2} \sim 10^5$~cm$^{-3}$ \citep{Lockett1999}. This theoretical prediction was confirmed observationally in several SNRs \citep[e.g.][]{Frail1998, Brogan2013}. 

 %%%%%%%%%%%%%%%%%%%%% FIGURE 3 %%%%%%%%%%%%%%%%%%%%%%%%%%%%%%%%%%%
\begin{figure*}
\centering
    \includegraphics[width=0.9\textwidth]{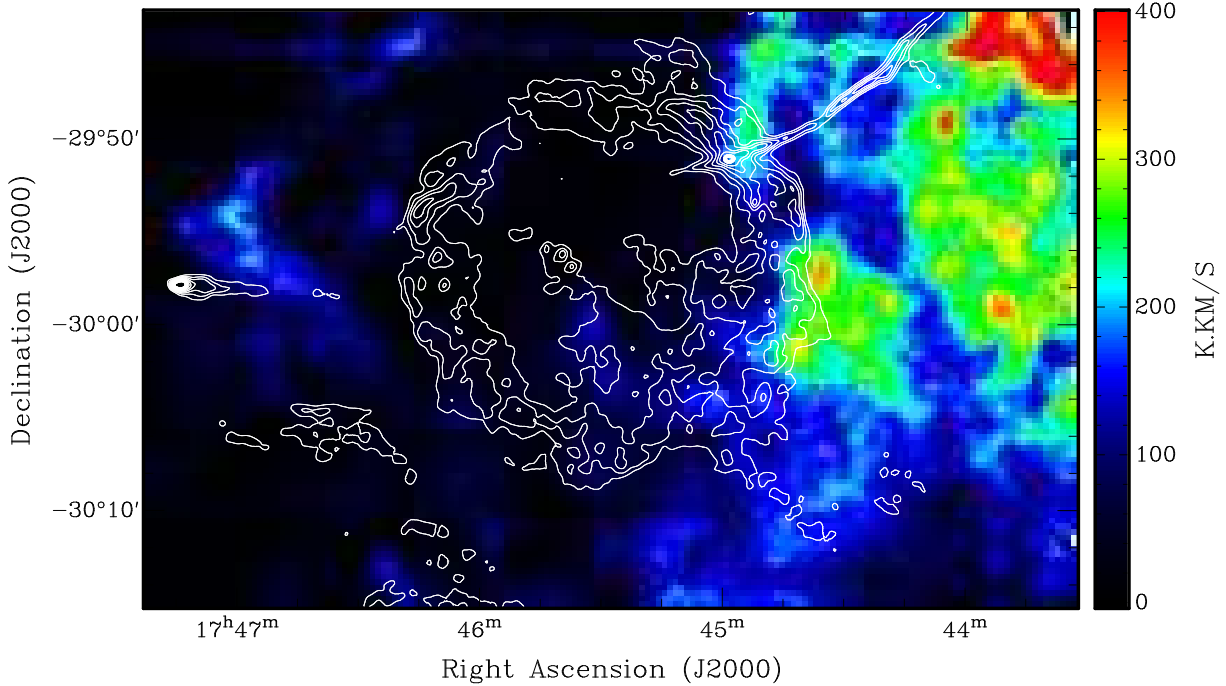}
    \caption{
    Distribution of $^{12}$CO emission integrated between $-178$ and\-  $-60$~km~s$^{-1}$. The beam size is  53\hbox{$.\!\!{}^{\prime\prime}$}6~$\times$~53\hbox{$.\!\!{}^{\prime\prime}$}6. The color scale is indicated at the right in units of K~km~s$^{-1}$.  White contours at 0.0035, 0.008, 0.014, 0.019 and 0.025~mJy~beam$^{-1}$ represent the radio emission at 1.5~GHz.}
    \label{fig:anillo}
    \vspace{5pt}
\end{figure*}
%%%%%%%%%%%%%%%%%%%%%%%%%%%%%%%%%%%%%%%%%%%%%%%%%%%%%%%%%%%%%%

%%%%%%%%%%%%%%%%%%%%%%%%%%%%%%%%%% FIGURA 4: CO/HESS %%%%%%%%%%%%%%%%%%
\begin{figure*}
\centering
   \includegraphics[width=0.9\textwidth]{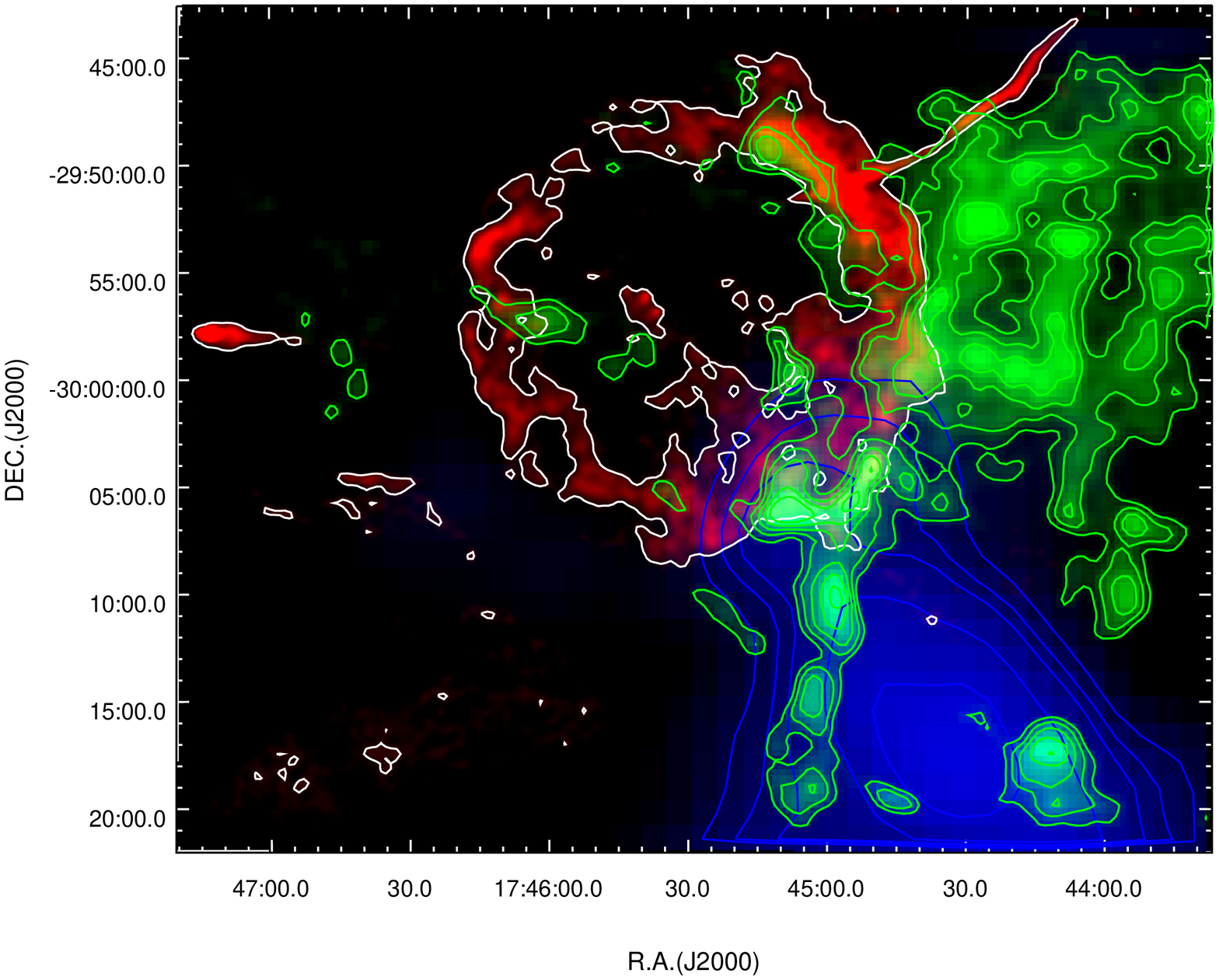}
    \caption{Composite RGB image of G359.1-0.5 and HESS J1745-303, where red is the radio emission at 1.5~GHz, green is the $^{12}$CO emission (see Fig. \ref{fig:filam}) and blue is the TeV $\gamma$-ray emission observed with HESS, with blue contours overlaid. Radio continuum white contours at 0.008, 0.019 and 0.025~mJy~beam$^{-1}$ are also included.}
    \label{fig:Hessyfilam}
    \vspace{2pt}
\end{figure*}

%%%%%%%%%%%%%%%%%%%%%%%%%%%%%%%%%%%%%%%%%%%%%

\cite{Lazendic2002} observed a $\sim$ 2\hbox{$^{\prime}$}$\times$2\hbox{$^{\prime}$} area around maser A in the H$_2$ 1-0 S(1) line and found a linear feature of 1\hbox{$.\!\!{}^{\prime}$}5-length and 15$^{\prime\prime}$-width containing the maser and lying parallel to the edge of the radio shell. They determined a significant density gradient between the eastern (10$^4$~cm$^{-3}$) and western (10$^5$~cm$^{-3}$) sides of this feature, supporting the idea that the H$_2$ originates in the expansion of the shock driven by G359.1-0.5. The H$_2$ linear feature tightly surrounds clump a$_2$ to the West (Fig.~\ref{fig:grumos}). Based on the relative position and difference in density, we propose that clump a$_2$ traces the pre-shock gas, while the western edge was compressed by the SNR shock front, attaining the physical conditions necessary to produce amplified maser emission at 1720 MHz. The linear feature was also observed in the CS J=3-2 line \citep{Lazendic2002}, from which an H$_2$ density of 10$^5$~cm$^{-3}$ was derived, in coincidence with the values obtained through the H$_2$ line. 

As mentioned in Sect.~\ref{Intro}, \cite{YusefZadeh1995}  also found an elongated, weaker maser component that extends about 5$^{\prime}$ and appears to link masers A and B. This OH feature spans from $-11$ to $-3$~km~s$^{-1}$, with a brightness peak at about $-8$ km~s$^{-1}$. A comparison with Fig.~\ref{fig:grumos} reveals that the maser emission is placed to the West of the CO emission and stretches beyond the limits of clump a$_1$. Extended OH (1720 MHz) maser emission is a powerful tracer of shocked gas \citep{Hewitt2008}.

\subsection{CO distribution at other velocities}
\label{sec:radio2}

As mentioned before, \citet{Uchida1992b} described a $^{12}$CO structure detected in a 1\hbox{$.\!\!{}^\circ$}5 $\times$ 1\hbox{$.\!\!{}^\circ$}5 field 
surrounding G359.1-0.5 with an annular morphology concentric with the radio shell and within the velocity range from $-190$ to $-60$ km s$^{-1}$. After the discovery of the five OH (1720~MHz) masers, which provides a strong indication of the systematic velocity of the clouds interacting with SNRs \citep{Frail1996}, the association between the CO ring reported by \cite{Uchida1992b} and G359.1-0.1 can be treated as coincidental. This is not too surprising considering that both molecular clouds and SNRs tend to be circular in shape, and chance alignments are very likely. Nevertheless, to investigate the distribution of the clouds at those velocities towards this region, we inspected our $^{12}$CO and $^{13}$CO data. Fig.~\ref{fig:anillo} shows the CO emission integrated from $-60$ up to the spectral limit of our observations, -178~km~s$^{-1}$. We observe a structure to the West partially overlapping the radio continuum shell, but we find no evidence of the molecular ring at these velocities. 

%%%%%%%%%%%%%%%%%%%%%%%%%%%%%%%%%% FIGURA 5: CO/HESS %%%%%%%%%%%%%%%%%%
\begin{figure*}
   \includegraphics[width=0.9\textwidth]{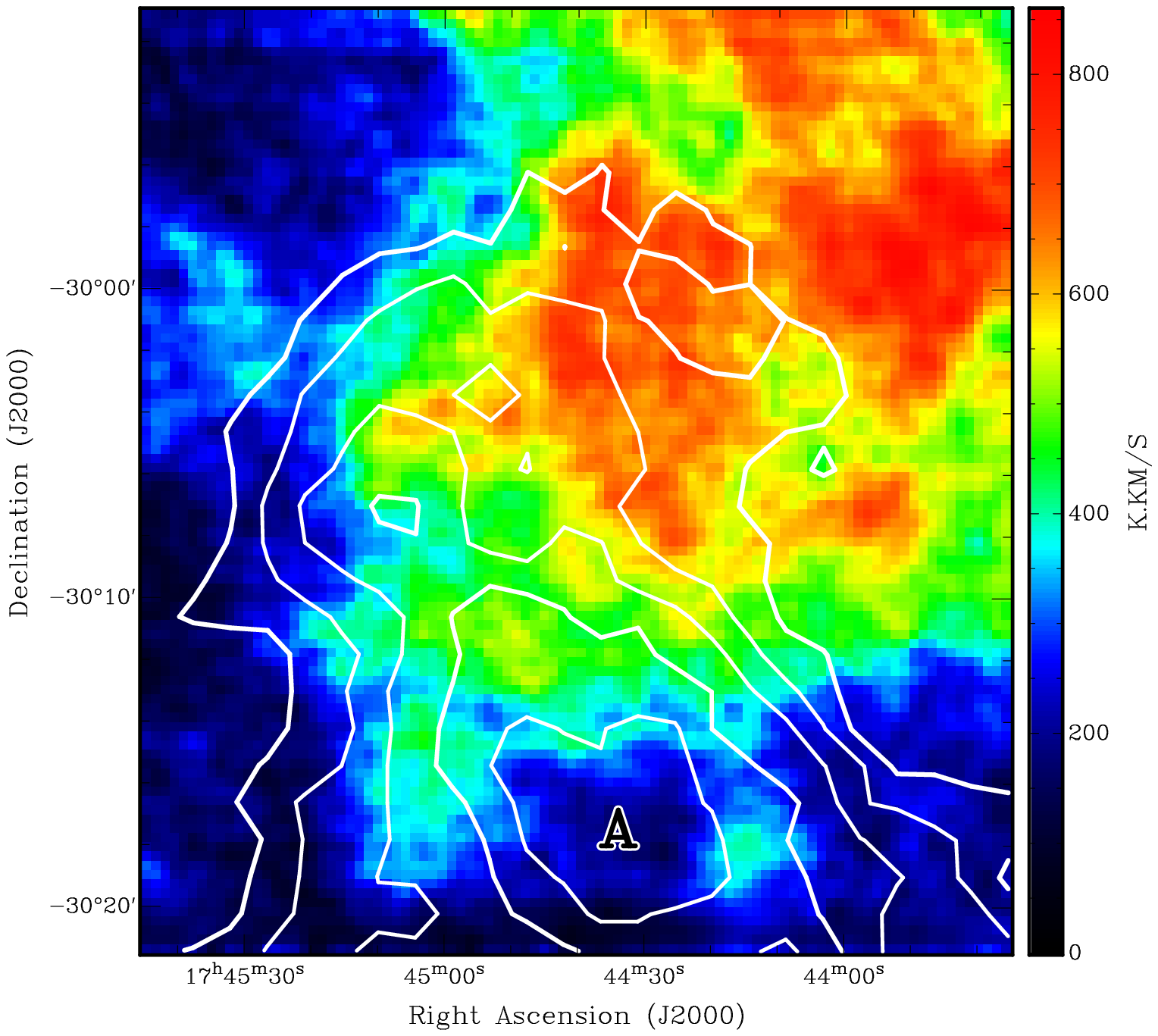}
    \caption{Distribution of the $^{12}$CO emission in the vicinity of HESS J1745-303's region A integrated between -170 and +160~km~seg$^{-1}$. The colour scale intensity is shown at the right. The TeV $\gamma$-ray emission is represented in white contours. The  $\gamma$-ray peak A is coincident with a CO minimum. 
    }
    \label{fig:HessyCO}
    \vspace{1pt}
\end{figure*}
%%%%%%%%%%%%%%%%%%%%%%%%%%%%%%%%%%%%%%%%%%%%%%%%%%%%%%%%%%%%%%%%%%%%%

\subsection{Comparison between CO distribution and emission at $\gamma$-rays}

The $\gamma$-ray source HESS J1745-303, adjacent to G359.1-0.5, was first discovered in the High Energy Stereoscopic System (H.E.S.S.) Galactic Plane Survey \citep{Aharonian2006}. This extended TeV complex has been described as consisting of three components \citep{Aharonian2008}. There is growing evidence that region A, the component closer to G359.1-0.5, is unrelated to the other two components \citep{Hui2011, Hui2016, Wilt2017}, hence we will focus only on this one. In Fig.~\ref{fig:Hessyfilam}, the TeV emission is represented in blue, while the radio continuum emission is depicted in red and highlighted with white contours, and the CO emission integrated between $-12.48$ and $+1.83$~km~s$^{-1}$ (see Fig.~\ref{fig:filam}) is represented in green. 

Very high energy (VHE) $\gamma$-rays are basically produced by three mechanisms: (i) up-scatter of lower-energy to higher-energy photons by interactions with high-energy electrons via the inverse-Compton process, (ii) acceleration of electrons in the electrostatic fields of ions and atomic nuclei via relativistic Bremsstrahlung, and (iii) $\pi^0 -$ decay via interactions of high-energy hadrons (cosmic rays) with the ambient gas. Although the lack of a spectral break in HESS J1745-303 is not compatible with many other GeV-SNRs \citep{Acero2015}, the total energy of cosmic-ray protons accelerated in G359.1-0.5 can produce the observed the TeV emission  \citep[e.g.][]{Hui2016}. Therefore, several previous works have tried to explain HESS J1745-303 in terms of the SNR through the hadronic mechanism. 

To explore the connection with molecular material, \cite{Aharonian2008} made use of a CO survey towards the Galactic Center performed by \citet{Bitran1997} and obtained a map integrated between $-100$ and $-60$~km~s$^{-1}$. We note that this survey is undersampled and has an HPBW of 8\hbox{$.\!\!{}^{\prime}$}8, much larger than the $100^{\prime\prime}$ beam of the CO data in \citet{Uchida1992b}. However, they have disregarded the fact that these velocities do not match those of the masers found around the SNR, and a cloud was found partially overlapping HESS J1745-303 from which the authors estimate that the interaction between G359.1-0.5 and a molecular cloud can account for the $\gamma-$rays emission in 'region A'. 

\citet{Hayakawa2012} observed an area covering G359.1-0.5 and HESS J1745-303 in the CO J:2-1 line as part of the NANTEN2 Galactic-Center Survey. They detected the same cloud as \citet{Aharonian2008} but integrating in a wider velocity range, up to $-15$~km~s$^{-1}$, and with a much better resolution (HPBW=$60^{\prime\prime}$). The full area was also surveyed by \citet{Wilt2017} in HC$_3$N, H$_2$O and several NH$_3$ transitions with the Mopra radio telescope and FWHM of $\sim$2~arcmin at a wavelength of 12~mm. Their maps are integrated between $-200$ and $+200$~km~s$^{-1}$ and show patchy molecular emission in the vicinity of the west-south side of G359.1-0.5. In particular, they report spectral broadenings of more than 10~km~s$^{-1}$ at 'region A' (acknowledging that the kinematic velocities differ from the masers velocity by $\sim$50~km~s$^{-1}$) and an enhanced ratio of NH$_3$ (3,3) over NH$_3$ (2,2), all indicative of heated gas. The authors attribute such heating to the passage of a shock front of unknown origin, since the clouds are outside the edge of the SNR. 

Using the systematic velocity of the molecular gas associated with G359.1-0.5, based on the maser velocities, we produced an integrated CO image in Fig.~\ref{fig:Hessyfilam} (see also Fig.~\ref{fig:filam}). The velocity range encompassing this structure differs by at least 50~km~s$^{-1}$ from previous molecular studies towards HESS J1745-303, posing an utterly diverging picture when the new CO distribution and the TeV emission are compared. We observe in Fig.~\ref{fig:Hessyfilam} an overlap between the southern tail of the CO structure, the SNR shell at the SW and the northern end of HESS J1745-303. If we are to judge solely on morphological arguments, this spatial configuration could suggest that G359.1-0.5 is possibly triggering the observed $\gamma$-ray emission through hadronic interactions.
However, there is no correlation between the CO distribution and the TeV intensity. On the contrary, the TeV peak appears to be surrounded, not overlapped, by a chain of molecular cloudlets.

In addition, we inspected the CO data cubes throughout the whole spectral ranges observed and found no positional coincidence between the CO emission and the $\gamma$-ray peak at any velocity. Moreover, the CO distribution unveils a minimum exactly coincident with the position of the TeV emission peak, as can be observed in the image integrated between $-170$ and $+160$~km~s$^{-1}$ displayed in Figure~\ref{fig:HessyCO}. With the absence of a correlation between the CO and the $\gamma$-ray emission at any velocity from our data, the nature of HESS J1745-303 remains a mystery but can hardly be reconciled with a hadronic origin connected with G359.1-0.5.

\section{Conclusions}
\label{sec:conclu}

We have carried out a molecular survey using the $^{12}$CO and $^{13}$CO J:$1-0$ lines towards the SNR G359.1-0.5. This study is complemented with archival HESS maps towards this region. The main findings can be summarized as follows:

\begin{enumerate}
  \item Four molecular clumps with densities $\sim 10^3$~cm$^{-3}$, some of them following closely the curvature of the remnant, have been detected in projection onto OH (1720~MHz) masers and at their same velocities.
 
  \item On this basis, we identified a larger-scale molecular structure between $v_{\rm LSR} =-12.48$ and $+1.86$~km~s$^{-1}$ extending to the West of G359.1-0.5. We propose that this structure, as well as the cloudlets associated with the OH (1720~MHz) masers, depict the pre-shock molecular gas into which the SNR shock front is expanding.
  
  \item We think it is highly unreliable that the ISM structures reported in the literature at velocities exceedingly different from those of the OH (1720~MHz) masers are  related to the SNR.
  
  \item The molecular cloud distribution at the masers velocities poorly matches the $\gamma$-ray distribution of HESS J1745-303. Thus, the $\gamma$-rays are unlikely to be directly related to the SNR. Inspection of our spectral data cubes at other velocities also does not show any molecular feature correlated with the intensity distribution of the $\gamma$-ray emission.

\end{enumerate}

Future studies using molecular species tracers of high density gas, like CS, C$^{18}$O, or H$_2$CO \citep{Mangum1993,Bronfman1996}, would be very helpful to identify shock signatures at possible interaction sites, particularly around masers D and E where no CO-maser correlation could be established.  

\section*{Acknowledgements}
LKE is supported by a CONICET fellowship. EMR, JAC and JFAC are members of the Carrera del Investigador Cient\'\i fico of CONICET, Argentina. This research was partially funded by CONICET grants PIP 112-201207-00226 and 112-201701-00604. 
JAC and JFAC. were supported by PIP 0102 (CONICET) and PICT-2017-2865 (ANPCyT). This work was also supported by the Agencia Estatal de Investigaci\'on grant AYA2016-76012-C3-3-P from the Spanish Ministerio de Econom\'\i a y Competitividad (MINECO) and by the Consejer\'\i a de Econom\'\i a, Innovaci\'on, Ciencia y Empleo of Junta de Andaluc\'\i a under research group FQM-322, as well as FEDER funds.
%%%%%%%%%%%%%%%%%%%%%%%%%%%%%%%%%%%%%%%%%%%%%%%%%%

%%%%%%%%%%%%%%%%%%%% REFERENCES %%%%%%%%%%%%%%%%%%

% The best way to enter references is to use BibTeX:

\bibliographystyle{mnras}
\bibliography{G359} % if your bibtex file is called example.bib

% Alternatively you could enter them by hand, like this:
% This method is tedious and prone to error if you have lots of references
%\begin{thebibliography}{99}
%\end{thebibliography}

%%%%%%%%%%%%%%%%%%%%%%%%%%%%%%%%%%%%%%%%%%%%%%%%%%

%%%%%%%%%%%%%%%%% APPENDICES %%%%%%%%%%%%%%%%%%%%%

%\appendix

%\section{Some extra material}

%%%%%%%%%%%%%%%%%%%%%%%%%%%%%%%%%%%%%%%%%%%%%%%%%%

% Don't change these lines
\bsp	% typesetting comment
\label{lastpage}
\end{document}